\documentclass[a4paper,twocolumn,english,aps,prl]{revtex4}
\usepackage[T1]{fontenc}
\usepackage[latin1]{inputenc}
\usepackage{float}
\usepackage{graphicx,color}

\makeatletter

\usepackage{babel}
\makeatother
\begin{document}

\preprint{This line only printed with preprint option}

\title{Optimal electron entangler and single electron source at low temperatures}

\author{Y. Sherkunov}

\affiliation{Department of Physics, University of Warwick, Coventry, CV4 7AL,
UK}

\author{Jin Zhang}

\affiliation{Department of Physics, University of Warwick, Coventry, CV4 7AL,
UK}
\author{N. d'Ambrumenil}

\affiliation{Department of Physics, University of Warwick, Coventry, CV4 7AL,
UK}

\author{B. Muzykantskii}

\affiliation{Department of Physics, University of Warwick, Coventry, CV4 7AL,
UK}


\begin{abstract}
Electron transport in mesoscopic contacts at low temperatures is accompanied by 
logarithmically divergent equilibrium noise. We show that this equilibrium noise 
can be dramatically suppressed in the case of a tunnel junction with modulated 
(time-dependent) transparency, and identify the optimal protocol. 
We show how such a contact could be used either as an optimal electron 
entangler or as a single-electron source with suppressed
equilibrium noise at low temperatures.

\end{abstract}
\maketitle


The controlled production, manipulation and detection of entangled particles are central to 
quantum computation. Considerable progress has been made on the manipulation 
of several qubits in quantum optics \cite{Kok07}. 
Given the controllability of device parameters 
and the rapid and coherent transport through mesoscopic contacts, one 
expects that electrons
could also be used to process quantum information  \cite{Beenakker05}. 
Although the 
entanglement of electron quasiparticles has not been demonstrated experimentally, 
important steps towards the detection of entangled states in mesoscopic systems have been taken 
\cite{NederNature} and there are many theoretical proposals on how to generate, manipulate and 
detect the entangled electrons in mesoscopic systems \cite{Beenakker05,Samuelsson05, Beenakker04, 
Trauzettel06, BeenakkerTitov05, Lebedev04, Samuelsson04, Lorenzo05}. 

Another breakthrough in `electronic optics' is the experimental 
demonstration of a single-electron source on demand with sub-nanosecond time 
resolution \cite{Feve07}. F\`{e}ve \textit{et.al.} used the coupling of the 
localized level of a quantum dot to a ballistic conductor to obtain a sequence 
of single-electron pulses \cite{Feve07}. 

To manipulate entangled electrons in mesoscopic systems requires low electron temperatures,
$T$ 
\cite{Buttiker08}. However, when $Tt_f \ll 1$ ($t_f$ is the measurement time), the problem of 
equilibrium noise associated with the opening and closing of channels
arises \cite{LesovikLeeLevitov96}. If the connection-disconnection process is abrupt, this leads to 
a logarithmically divergent (so-called) equilibrium noise 
$\langle\langle Q^2\rangle \rangle\propto \log t_f/\tau_s$, where $\tau_s$ is the 
characteristic switching time \cite{LesovikLeeLevitov96}
(see also \cite{Moskalets07}).

Here, we show how the appropriate choice of the time-dependence of the barrier transparency in a 
tunnel junction can dramatically reduce the equilibrium noise. We suggest 
how such point contacts could be used as sources for  
entangled quasiparticles and ultracold single electrons.
Firstly, we describe
a source of entangled particles, based on two 
one-dimensional wires coupled by a tunnel
junction with tunable transmission and reflection amplitudes $A$ and $B$, 
which attains the theoretical maximum entanglement 
of 50\% \cite{BeenakkerTitov05}. 
We show that tuning the transparency amplitude to take the time-dependent form
$A=\mbox{Im} \prod_j^N\frac{t-t_j-i\tau_j}{t-t_j+i\tau_j}$ 
($\tau_j>0$, or $\tau_j<0$ $\forall j$)  
leads to the excitation of $N$ independent spin-entangled particle-hole pairs, with 
finite $\tau_j$-independent noise. In the special case $N=1$, the particle and 
hole are distributed independently between two leads with probability $1/2$ 
to find the 
particle (or the hole) in either lead. Our proposed scheme is similar to the one 
based on the quantum 
pump \cite{BeenakkerTitov05} but with no logarithmically divergent 
equilibrium noise.  

Secondly, we propose a single-electron source, based on a biased quantum point contact 
with tunable transparency. It has been shown \cite{IvaLL97,KeelingKlichLev06}, that 
a Lorentzian voltage pulse $V(t)$ applied between the leads of a quantum point contact 
excites a single electron (or hole), provided the Faraday flux,  
$\psi=e/\hbar \int^t V(t')dt'$, is an integer multiple of $2 \pi$. Even at low temperatures the 
transport of a single electron is again accompanied by the logarithmically divergent equilibrium 
noise, which makes detection of the excitation difficult. We show that the equilibrium 
noise in the scheme proposed in \cite{IvaLL97} can be suppressed by tuning the barrier 
transparency to be $A=\mbox{Im} \frac{(t+t_1/2-i\tau_1)(t-t_1/2+i\tau_1)}{(t+t_1/2+i\tau_1)(t-t_1/2-i\tau_1)}$. 
If a quantized Lorentzian pulse is applied between the leads when the barrier is open, 
the tunnel junction will operate as a very low noise single-electron 
source. 

The noise in the charge transferred across
a point contact and the degree of entanglement of states on either side of the contact
can be controlled by the time-dependence of the barrier transparency. 
The noise in the charge transferred across the barrier 
will be given by
$\langle \langle Q^2 \rangle \rangle= \int dt dt' 
[2A(t)^2A(t')^2 n_L(1-n_R)+A(t)B(t)A(t')B(t')(n_L(1-n_R)+(1-n_L)n_R)]$, 
where $A(t)$ and $B(t)$ are the (time-dependent) transmission and 
reflection amplitudes of the barrier, $n_{L(R)}$ is the density matrix of incoming states 
in the left (right) lead \cite{LeeLevitov93}. 
With no bias voltage,
$n_{L,R}=n_0=i/[2\pi(t-t'+i0)]$.
If the transparency 
amplitude $A$ switches abruptly, $A\neq 0$ for $t\in [0,t_f]$ and $A=0$ otherwise, 
one obtains the well-known logarithmically divergent
result for the zero bias ($V=0$) noise 
$\langle \langle Q^2 \rangle \rangle_{_0} =\frac{A^2}{\pi^2}\log t_f/\tau_s$.
This logarithmic divergence
is an unavoidable consequence of the abrupt start and finish to the measurement
and simply reflects the density fluctuations of fermions 
in 1D crossing any point during a time-interval $[0,t_f]$. 
If, instead, we choose $A=\mbox{Im} \frac{t-i\tau}{t+i\tau}$, 
we obtain 
$\langle \langle Q^2 \rangle \rangle_{_0} =1/2$. As we show below, this result
could be used together with the spin degree of freedom of electrons to define
an optimum entanglement protocol.

By combining a smooth opening of the contact and the
application of a voltage bias between the leads, we show that it is possible to 
operate the contact as a single electron source with high signal-to-noise ratio.
We propose tuning the barrier height so that
$A(t)=\mbox{Im}\frac{(t+t_1/2-i\tau_1)(t-t_1/2+i\tau_1)}{(t+t_1/2+i\tau_1)(t-t_1/2-i\tau_1)}$,
with $\tau_1=t_1(1/2+1/\sqrt{2})$, which ensures that the transparency amplitude 
goes through a single maximum at $t=0$ with $A(0)=1$. 
If a quantized bias voltage pulse 
is applied between the leads,  $n_L=n_0$, $n_R =e^{i\psi}n_0 e^{-i\psi}$ with
$e^{i\psi}=\frac{t-t_0-i\tau_0}{t-t_0+i\tau_0}$, which is narrow in time
($\tau_0 \ll \tau_1$),
we should expect
$\langle \langle Q^2 \rangle \rangle \approx 
\langle \langle  Q^2 \rangle \rangle_{_0}
+ |A(t_0)|^2(1-|A(t_0)|^2)$. The first term describes the zero bias 
noise associated with the opening of the contact 
and the second gives the shot noise for an open contact \cite{LesovikLeeLevitov96}.
For $t_0=0$ the second term can be made to vanish so that for this choice 
of $\tau_1/t_1$, we should  expect that  $\langle \langle Q^2 \rangle \rangle \approx
\langle \langle Q^2 \rangle \rangle_{_0} \approx 1/4$.

The statistics of the transferred charge are encoded in the characteristic
function, $\chi(\lambda)$ \cite{LevitovJETP93}:
\begin{eqnarray}
 \chi(\lambda)=\sum_{n=-\infty}^{+\infty}P_n e^{in \lambda}.
\label{chi}
\end{eqnarray} Here $P_n$ is the probability of $n$ particles being transmitted across 
a barrier. A variety of approaches have been used to study $\chi(\lambda)$, with the majority focused on 
voltage-biased contacts \cite{LevitovJETP93,IvaLL97,MA03,dAMuz05,VanevicNazBelz07,Nazarov08,
sherkunov08}. We find a mapping between the problem of the biased leads
and that of the time-dependent barrier transmission. Our formulation of the problem
enables us to solve for  $\chi(\lambda)$ analytically in an unbiased contact  and 
to compute numerically the characteristic function for  a
time-dependent transmission amplitude in the presence of bias voltages.

The system we consider is a 2-terminal mesoscopic contact at zero temperature, which 
consists of two single-channel leads and a barrier characterised by transmission 
and reflection  amplitudes,  $A$ and $B$. The annihilation operators of the incoming states  
$a_{L(R)}$ are related to the outgoing states $b_{L(R)}$ at the left (right) lead via:
\begin{equation} \left(\begin{array}{ll} b_L \\  b_R\end{array}\right)= S 
\left(\begin{array}{ll} a_L \\  a_R\end{array}\right) \hspace{0.2cm} 
\mbox{with} \hspace{0.2cm} S=\left( \begin{array}{ccc}
B & A \\
-A^* & B^* \end{array} \right), 
\label{S}
\end{equation} 
where the scattering matrix $S$ varies slowly 
compared to the Wigner delay time. 
We will take $A$ and $B$ to be real. Then, 
in the (time-independent) basis in which
the scattering matrix is diagonal, 
the incoming states
 $\tilde{a}_1=\frac{1}{\sqrt{2}}(a_L-ia_R)$ and $\tilde{a}_2=\frac{1}{\sqrt{2}}(a_L+ia_R)$.
In this basis 
$S_{11}=e^{i\phi(t)}$, $S_{22}=e^{-i\phi(t)}$, $S_{12}=S_{21}=0$, 
where $e^{i\phi(t)}=B(t)+iA(t)$, and
the outgoing states are 
$b_L=\frac{1}{\sqrt{2}}\left(\tilde{a}_1e^{i\phi}+\tilde{a}_2e^{-i\phi}\right)$ and 
$b_R=\frac{i}{\sqrt{2}}\left(\tilde{a}_1e^{i\phi}-\tilde{a}_2e^{-i\phi}\right)$.

\begin{figure}[!]
\centering
 \includegraphics[width=0.45\textwidth]{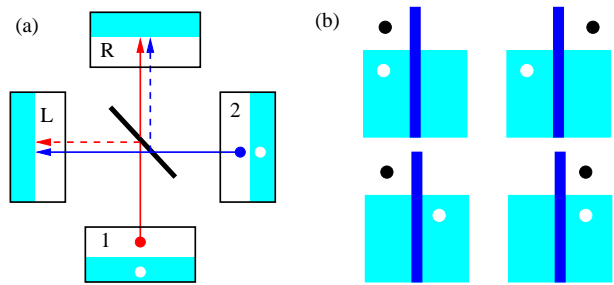}
\caption{\label{fig:fig11}(color online) Charge transport through an unbiased contact 
with time-dependent scattering matrix. (a) Particle-hole excitations induced
in channels $1$ and $2$ (in which the scattering matrix is diagonal) can be thought
of as propagating through a 50\% beam-splitter to yield the charge in the left ($L$) and
right leads ($R$) of the contact.
%
%
(b)
Outcomes of applying the pulse (\ref{MES}) to the barrier 
of the tunnel junction.   A hole (white) and an electron (black) occupies states 
in both leads with probability $1/2$. }
\end{figure}

The relation between the channels 1 and 2 and those of the two leads
$L$ and $R$ is the electronic equivalent of a beam splitter. This is illustrated in Fig.\ref{fig:fig11}a
for the case of single excitations in channels 1 and 2 .
The time-dependence of the barrier excites a particle-hole pair in channel 1 with probability 
$\sin^2\frac{\alpha_1}{2}$ and another in channel 2 with probability 
$\sin^2 \frac{\alpha_2}{2}$. 
For the unbiased case,
reflection and transmission of 
particles in channels 1 and 2 occur independently.
The probability of finding no additional charge in the right lead as a result of the excitation
in channel $j$ is then
$P_0=\cos^2 \frac{\alpha_j}{2} +\frac{1}{2}\sin^2 \frac{\alpha_j}{2}$, while the probability of finding 
an additional particle or hole are: $P_{\pm 1}=\frac{1}{4}\sin^2 \frac{\alpha_j}{2}$. 
The characteristic 
function $\chi(\lambda)$ is the product of the results from the two independent channels:
\begin{equation}
 \chi(\lambda)=\prod_{j}\left (1+\frac{1}{4}\sin^2 \frac{\alpha_j}{2}
\left(e^{i\lambda}+e^{-i\lambda}-2\right) \right).
\label{chi11}
\label{rez}
\end{equation}
The angles $\alpha_j$ can be found by diagonalizing the matrix $h_1h_2$ 
\cite{VanevicNazBelz07,sherkunov08}:
\begin{eqnarray}
 h_1h_2=e^{i\phi}he^{-i\phi}e^{-i\phi}he^{i\phi},
\label{h1h2}
\end{eqnarray}
where $h=2n_0-1$ is related to the density matrix, $n_0$,  of the ground state. The 
eigenvalues of $h_1h_2$ are paired and equal to $e^{i\alpha},e^{-i\alpha}$ 
\cite{VanevicNazBelz07,sherkunov08}. In the case of a voltage-biased contact, there can also be
unpaired eigenvalues, corresponding to so-called unidirectional events, in which a 
single particle or hole is excited. 

Eq.(\ref{rez}) formally coincides with the one obtained for
$\chi(\lambda)$ for bidirectional events in a quantum contact. For a (zero-mean) time-dependent
bias, $V(t)$, between the leads and a  barrier which is described 
by a time-independent scattering matrix \cite{VanevicNazBelz07}:
$ \chi(\lambda)=\prod_j\left (1+ |A|^2 |B|^2 \sin^2 \frac{\alpha_j}{2}
\left(e^{i\lambda}+e^{-i\lambda}-2\right) \right)$.  Setting 
$|A|^2=|B|^2=\frac{1}{2}$, gives the result (\ref{chi11}). 
In the case of the biased contact the angles $\alpha_j$ are found by diagonalizing 
$e^{i \psi(t)}he^{-i\psi(t)}h$, where 
$\psi=e/\hbar \int ^t V(t')dt'$ is the Faraday flux. Direct comparison with Eq.(\ref{h1h2}) 
shows that the problem of the unbiased quantum contact with time-dependent scatterer 
is equivalent to the problem of a voltage biased quantum contact with time-independent 
scatterer if:  
$\psi\rightarrow 2\phi$, $|A|^2 \rightarrow 1/2$, $|B|^2 \rightarrow 1/2$. 

As an illustration of the power of this mapping,  we consider the case of a periodically 
modulated transparency $A=\sin(\omega t)$, $B=\cos(\omega t)$. The mapping means that
this  problem is equivalent to that of a contact under constant bias. 
The characteristic function for this case is known to be
$\chi(\lambda)=\left[\chi_0(\lambda)\right]^{\frac{t_f \omega}{2\pi}}$ \cite{LevitovJETP93},
where $\chi_0$ describes a single period, and $t_f$ is the measurement time. The 
eigenvalues of the matrix $he^{2i\omega t}he^{-2i\omega t'}$ for a single period are 
paired and equal to $-1$. Using this to compute $\chi(\lambda)$ for a barrier with
periodically modulated transparency gives:
$\chi(\lambda)=\left[(1+\cos\lambda)/2\right]^{\omega t_f/2\pi}$ which is a
result previously obtained by Andreev and Kamenev \cite{Andreev-Kamenev2000}.

The problem of voltage-biased quantum point contacts with constant transparency 
amplitude $A$ , where a voltage $V(t)$ is applied to one of the leads of a tunnel
junction has been extensively studied \cite{LevitovJETP93, IvaLL97, LesovikLeeLevitov96,
VanevicNazBelz07, sherkunov08}. 
Particular attention has been paid to the case of quantized Lorentzian voltage 
pulses applied to one electrode $V(t)=\sum_j^N\frac{-2\tau_j}{(t-t_j)^2+\tau_j^2}$ 
($\tau_j>0$ or $\tau_j<0$ $\forall$ $j$) when
\begin{eqnarray}
 e^{i \psi(t)}=\prod_{j=1}^N\frac{t-t_j-i\tau_j}{t-t_j+i\tau_j}.
\label{phi1}
\end{eqnarray}
If 
$\tau_j>0$ or $\tau_j<0$ 
the pulse excites exactly $N$ particles or holes depending on the polarity of the device
\cite{IvaLL97,KeelingKlichLev06}.

Some of the results obtained for the problem of the biased point-contact,
when mapped to the time-dependent barrier case,
help with the design of an optimum entangler of particle states and the design of
low noise sources of cold electrons and holes. 
From (\ref{phi1}), we can see the special status of events in which the barrier is modulated
according to 
\begin{eqnarray}
 e^{i\phi}=B+iA=\sqrt{\left(\frac{t-i\tau}{t+i\tau}\right)
\left(\frac{t-t_1-i\tau}{t-t_1+i\tau}\right)}.
\label{MES}
\end{eqnarray}
(In the case $t_1=0$, Eq.(\ref{MES}) describes a single 
pulse of the type (\ref{phi1}).) The characteristic function, 
$\chi(\lambda)$, is found from the eigenvalues of $h_1h_2$ (see Eq.(\ref{h1h2})), which 
are the same as those of $he^{2i\phi}he^{-2i\phi}$. In the language 
of the biased lead case, these are just
two so-called uni-directional events and 
both eigenvalues are equal to -1, so $\alpha_{1,2}=\pi$ \cite{sherkunov08}.
From (\ref{rez}) we then obtain
$\chi(\lambda) = \frac{(1+\cos\lambda)}{2}.$
The four possible outcomes of applying the pulse (\ref{MES})
to the barrier are illustrated in Fig.\ref{fig:fig11}b 
and all occur with probability $1/4$. 

The particle and the hole in Fig.\ref{fig:fig11}b
are entangled. However, this entanglement cannot be 
revealed by measurements because of particle number conservation \cite{Beenakker04}. 
In order to create ``useful'' entangled states, which can be measured with the Bell 
procedure, we should consider entanglement in, for example, the spin degree of freedom
of the particles. 
Taking account of spin, each lead has two channels: one for each spin projection.
If we suppose that the scattering matrix for the barrier is spin-independent
(so that the two channels in each lead are independent), and a pulse
(\ref{MES}) is applied, 
we have $\chi(\lambda)=[\frac{1}{2} (1+\cos\lambda)]^2$. 
There are now $16$ outcomes which occur with equal probabilities.
The fully entangled Bell pair is 
created only if the change in charge in one of the leads is $+1$ and $-1$ in the other,
and occurs with
probability $1/2$. The measurement procedure 
requires measurement of the charge in one of the leads first, and then of 
the spin state of the particle or hole. 

To calculate the available entanglement 
entropy, one should use the super-selection rules, which
account for particle number conservation \cite{Wiseman03} (see also \cite{Klich081,Klich082}). Let a quantum state, with 
$N$ particles distributed between Alice and Bob (left and right leads), be
described by a wavefunction $|\Psi_{AB}\rangle$ and let $|\Psi_{AB}^n\rangle$ 
be the wavefunction projected onto a subspace of fixed local particle number $n$ for Alice 
and $N-n$ for Bob.  The available entropy is: 
$S_{avl}=\sum_n P_n S_n$, where $P_n$ is the probability 
$P_n=\langle\Psi_{AB}^n|\Psi_{AB}^n\rangle/\langle\Psi_{AB}|\Psi_{AB}\rangle$ and 
$S_n$ is the standard entanglement entropy corresponding to the configuration $n$.
The probability to find a particle in the left lead and a hole in the right lead or 
vice versa is $1/4$. The entanglement entropy of the configurations 
containing a fully entangled Bell pair is $S_n=1$. Thus we find $S_{avl}=1/2$. 
This corresponds to an entangler with 50\% efficiency, which is the optimal value. 
The entangler proposed originally in \cite{BeenakkerTitov05} has the same efficiency
but still contains the equilibrium noise, which in our proposal is suppressed. 
Finally we note that for the case of $N$ pulses applied to the barrier 
$e^{i\phi}=\prod_j^N\sqrt{\left(\frac{t-t_j-i\tau_j}{t-t_j+i\tau_j}\right)
\left(\frac{t-t_j'-i\tau_j}{t-t_j'+i\tau_j}\right)}$ ($\tau_j>0$ or 
$\tau_j<0$ $\forall$ $j$) we find $\chi(\lambda)=\left(\frac{1+\cos\lambda}{2}\right)^N$,
{\it i.e.\/} the particle-hole pairs are produced independently.


\begin{figure}[!]
\includegraphics[width=0.45\textwidth, height=0.2\textheight]{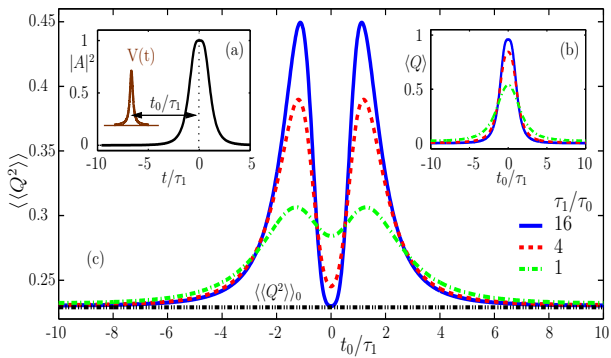}
\caption{\label{fig:fig3} (color online) Single unidirectional event in a quantum contact 
with tunable transparency.   (a) Transparency of the 
barrier, $|A(t)|^2$, as a function of time (see \ref{barr}). Also shown is the quantized 
voltage pulse, $V(t)$, applied to the left lead; (b) Average charge transfer  through 
the barrier, $\langle Q \rangle$, as a function of $t_0/\tau_1$, where $t_0$ is the separation between
the maxima of $V(t)$ and $|A(t)|^2$, see Fig.
\ref{fig:fig3}(a). The maximum of $\langle Q \rangle$ occurs when $t_0=0$ where
$|A|^2$ approaches 1, it grows as the width of 
the voltage pulse, $\tau_0$, is reduced, and is close to $1$ 
for $\tau_0=\tau_1/16$. (c) Noise, $\langle \langle Q^2 \rangle \rangle$, 
in a biased contact with transparency given by 
(\ref{barr}) as a function of $t_0/\tau_1$. The minimum of $\langle \langle Q^2 \rangle \rangle$
occurs for $t_0=0$ and, for a narrow voltage pulse ($\tau_0=\tau_1/16$), 
it approaches the limit set by the barrier modulation: 
$\langle \langle Q^2 \rangle \rangle_{_0}$ (horizontal dashed line). The maxima in the noise 
occur when $t_0/\tau_1 \approx 1$ and $|A|^2 \approx 1/2$.}
\end{figure}

Now we turn our attention to the ultracold single-electron source. 
We consider a tunnel junction with tunable transparency, 
which connects 
two single-channel wires.  A single quantized voltage 
pulse, with  $e^{i\psi}=\frac{t-t_0-i\tau_0}{t-t_0+i\tau_0}$, is applied to one of the 
leads, and a single particle is excited  \cite{IvaLL97}. If the barrier transparency 
is turned on abruptly, the logarithmically divergent equilibrium noise emerges
and makes measurement of a single unidirectional event problematic. To suppress 
the equilibrium noise, we propose modulating the barrier smoothly, by applying 
a pulse sequence to the barrier with: 
\begin{eqnarray}
e^{i\phi}=B+iA=\left(\frac{t+t_1/2-i\tau_1}{t+t_1/2+i\tau_1}\right)
\left(\frac{t-t_1/2+i\tau_1}{t-t_1/2-i\tau_1}\right).
\label{barr}
\end{eqnarray}
We choose $\tau_1=t_1(\frac{1}{2}+\frac{1}{\sqrt{2}})$. This choice gives a transparency 
amplitude, which does not change sign and goes 
through a single maximum at $t=0$, with $|A(0)|^2=1$, as shown in Fig.\ref{fig:fig3}a. 

The characteristic function in the general case of quantum pumping can be 
expressed in terms of the eigenvalues of the single-particle density matrix of 
outgoing states in, say, the left lead 
$n_{out}=\langle b_L^{\dagger}b_L\rangle=B^*n_0 B+A^*\tilde{n}A$. Here  
$\tilde{n}=e^{i\psi} n_0 e^{-i\psi}$. 
Consider first the case of a single-particle excitation in the outgoing states. 
The eigenvalues of the density matrix in the left lead are $1$ (for the states below 
the Fermi level), $0$ (for the states above the Fermi level not occupied by the
excitation) and $n_j$ for the state 
affected by the excitation. The characteristic function (\ref{chi}) 
is then $\chi_p(\lambda)=1-n_j+e^{i\lambda}n_j$. For a single hole in the outgoing 
states we find $\chi_h(\lambda)=n_j+e^{-i\lambda}(1-n_j)$. These two formulae 
can be combined to give $\chi(\lambda)=e^{i(Q_1-n_j)\lambda}(1-n_j+e^{i\lambda}n_j)$. 
(The transferred charge $Q_1=n_j$ for the particle excitation and $Q_1=n_j-1$ 
for the hole excitation.) In the general case the density 
matrix is Hermitian and thus possesses an orthonormal eigenbasis, and we arrive 
at the  formula of \cite{Abanov08}: 
$\chi(\lambda)=e^{i\lambda Q} \prod_j e^{-i\lambda n_j} \left[1+(e^{i\lambda}-1)n_j\right]$
where $Q$ is the total transferred charge. 

For the case of the pulse (\ref{barr}) applied to the barrier together with the 
bias pulse $e^{i\psi}=\frac{t-t_0-i\tau_0}{t-t_0+i\tau_0}$ applied to the left lead,
we have diagonalized the density matrix $n_{out}$ numerically. 
In Fig.\ref{fig:fig3}c  
we present results for the noise $\langle \langle Q^2\rangle \rangle$ in the 
system as a function of the separation $t_0$ between the bias pulse 
and the barrier pulses.  
The maximum values of the noise occurs when $t_0 \approx \tau_1$ and the 
transparency coefficient is almost $1/2$. This is the regime where the barrier 
is acting as a $50\%$ beam-splitter. 
The minimum of $\langle \langle Q^2\rangle \rangle$ corresponds to $t_0=0$, 
when $|A|^2=1$.  At $t_0=0$
the transferred charge also reaches its maximum 
(see Fig.\ref{fig:fig3}b). If the bias pulse is narrow compared to the 
barrier pulse ($\tau_0 \ll \tau_1$), 
the major contribution to the noise at the minimum is very close
to the value expected on the basis of our earlier heuristic argument, namely 
$\langle \langle Q^2 \rangle \rangle_{_0}$
and the transferred charge approaches $1$. This is the regime which 
we are proposing could be used as a single-electron source on 
demand at ultralow temperatures. 

\begin{acknowledgements}
The work was supported by EPSRC-EP/D065135/1. 
\end{acknowledgements}

\bibliographystyle{apsrev}
\bibliography{fcs+entropy}

\begin{thebibliography}{28}
\expandafter\ifx\csname natexlab\endcsname\relax\def\natexlab#1{#1}\fi
\expandafter\ifx\csname bibnamefont\endcsname\relax
  \def\bibnamefont#1{#1}\fi
\expandafter\ifx\csname bibfnamefont\endcsname\relax
  \def\bibfnamefont#1{#1}\fi
\expandafter\ifx\csname citenamefont\endcsname\relax
  \def\citenamefont#1{#1}\fi
\expandafter\ifx\csname url\endcsname\relax
  \def\url#1{\texttt{#1}}\fi
\expandafter\ifx\csname urlprefix\endcsname\relax\def\urlprefix{URL }\fi
\providecommand{\bibinfo}[2]{#2}
\providecommand{\eprint}[2][]{\url{#2}}

\bibitem[{\citenamefont{Kok et~al.}(2007)\citenamefont{Kok, Munro, Nemoto,
  Ralph, Dowling, and Milburn}}]{Kok07}
\bibinfo{author}{\bibfnamefont{P.}~\bibnamefont{Kok}},
  \bibinfo{author}{\bibfnamefont{W.~J.} \bibnamefont{Munro}},
  \bibinfo{author}{\bibfnamefont{K.}~\bibnamefont{Nemoto}},
  \bibinfo{author}{\bibfnamefont{T.~C.} \bibnamefont{Ralph}},
  \bibinfo{author}{\bibfnamefont{J.~P.} \bibnamefont{Dowling}},
  \bibnamefont{and} \bibinfo{author}{\bibfnamefont{G.~J.}
  \bibnamefont{Milburn}}, \bibinfo{journal}{Rev. Mod. Phys.}
  \textbf{\bibinfo{volume}{79}}, \bibinfo{eid}{135} (\bibinfo{year}{2007}).

\bibitem[{\citenamefont{Beenakker}(2006)}]{Beenakker05}
\bibinfo{author}{\bibfnamefont{C.~W.~J.} \bibnamefont{Beenakker}}, in
  \emph{\bibinfo{booktitle}{Proc. Int. School Phys E. Fermi}}
  (\bibinfo{publisher}{IOS Press, Amsterdam}, \bibinfo{year}{2006}).

\bibitem[{\citenamefont{Neder et~al.}(2007)\citenamefont{Neder, Ofek, Chung,
  Heiblum, Mahalu, and Umansky}}]{NederNature}
\bibinfo{author}{\bibfnamefont{I.}~\bibnamefont{Neder}},
  \bibinfo{author}{\bibfnamefont{N.}~\bibnamefont{Ofek}},
  \bibinfo{author}{\bibfnamefont{Y.}~\bibnamefont{Chung}},
  \bibinfo{author}{\bibfnamefont{M.}~\bibnamefont{Heiblum}},
  \bibinfo{author}{\bibfnamefont{D.}~\bibnamefont{Mahalu}}, \bibnamefont{and}
  \bibinfo{author}{\bibfnamefont{V.}~\bibnamefont{Umansky}},
  \bibinfo{journal}{Nature} \textbf{\bibinfo{volume}{448}},
  \bibinfo{pages}{333} (\bibinfo{year}{2007}).

\bibitem[{\citenamefont{Samuelsson and B\"uttiker}(2005)}]{Samuelsson05}
\bibinfo{author}{\bibfnamefont{P.}~\bibnamefont{Samuelsson}} \bibnamefont{and}
  \bibinfo{author}{\bibfnamefont{M.}~\bibnamefont{B\"uttiker}},
  \bibinfo{journal}{Phys. Rev. B} \textbf{\bibinfo{volume}{71}},
  \bibinfo{pages}{245317} (\bibinfo{year}{2005}).

\bibitem[{\citenamefont{Beenakker et~al.}(2004)\citenamefont{Beenakker,
  DiVincenzo, Emary, and Kindermann}}]{Beenakker04}
\bibinfo{author}{\bibfnamefont{C.~W.~J.} \bibnamefont{Beenakker}},
  \bibinfo{author}{\bibfnamefont{D.~P.} \bibnamefont{DiVincenzo}},
  \bibinfo{author}{\bibfnamefont{C.}~\bibnamefont{Emary}}, \bibnamefont{and}
  \bibinfo{author}{\bibfnamefont{M.}~\bibnamefont{Kindermann}},
  \bibinfo{journal}{Phys. Rev. Lett.} \textbf{\bibinfo{volume}{93}},
  \bibinfo{pages}{020501} (\bibinfo{year}{2004}).

\bibitem[{\citenamefont{Trauzettel et~al.}(2006)\citenamefont{Trauzettel,
  Jordan, Beenakker, and B\"uttiker}}]{Trauzettel06}
\bibinfo{author}{\bibfnamefont{B.}~\bibnamefont{Trauzettel}},
  \bibinfo{author}{\bibfnamefont{A.~N.} \bibnamefont{Jordan}},
  \bibinfo{author}{\bibfnamefont{C.~W.~J.} \bibnamefont{Beenakker}},
  \bibnamefont{and}
  \bibinfo{author}{\bibfnamefont{M.}~\bibnamefont{B\"uttiker}},
  \bibinfo{journal}{Phys.Rev. B} \textbf{\bibinfo{volume}{73}},
  \bibinfo{pages}{235331} (\bibinfo{year}{2006}).

\bibitem[{\citenamefont{Beenakker et~al.}(2005)\citenamefont{Beenakker, Titov,
  and Trauzettel}}]{BeenakkerTitov05}
\bibinfo{author}{\bibfnamefont{C.~W.~J.} \bibnamefont{Beenakker}},
  \bibinfo{author}{\bibfnamefont{M.}~\bibnamefont{Titov}}, \bibnamefont{and}
  \bibinfo{author}{\bibfnamefont{B.}~\bibnamefont{Trauzettel}},
  \bibinfo{journal}{Phys. Rev. Lett.} \textbf{\bibinfo{volume}{94}},
  \bibinfo{pages}{186804} (\bibinfo{year}{2005}).

\bibitem[{\citenamefont{Lebedev et~al.}(2004)\citenamefont{Lebedev, Blatter,
  Beenakker, and Lesovik}}]{Lebedev04}
\bibinfo{author}{\bibfnamefont{A.~V.} \bibnamefont{Lebedev}},
  \bibinfo{author}{\bibfnamefont{G.}~\bibnamefont{Blatter}},
  \bibinfo{author}{\bibfnamefont{C.~W.~J.} \bibnamefont{Beenakker}},
  \bibnamefont{and} \bibinfo{author}{\bibfnamefont{G.~B.}
  \bibnamefont{Lesovik}}, \bibinfo{journal}{Phys. Rev. B}
  \textbf{\bibinfo{volume}{69}}, \bibinfo{eid}{235312} (\bibinfo{year}{2004}).

\bibitem[{\citenamefont{Samuelsson et~al.}(2004)\citenamefont{Samuelsson,
  Sukhorukov, and B\"uttiker}}]{Samuelsson04}
\bibinfo{author}{\bibfnamefont{P.}~\bibnamefont{Samuelsson}},
  \bibinfo{author}{\bibfnamefont{E.~V.} \bibnamefont{Sukhorukov}},
  \bibnamefont{and}
  \bibinfo{author}{\bibfnamefont{M.}~\bibnamefont{B\"uttiker}},
  \bibinfo{journal}{Phys. Rev. Lett.} \textbf{\bibinfo{volume}{92}},
  \bibinfo{pages}{026805} (\bibinfo{year}{2004}).

\bibitem[{\citenamefont{Lorenzo and Nazarov}(2005)}]{Lorenzo05}
\bibinfo{author}{\bibfnamefont{A.~D.} \bibnamefont{Lorenzo}} \bibnamefont{and}
  \bibinfo{author}{\bibfnamefont{Y.~V.} \bibnamefont{Nazarov}},
  \bibinfo{journal}{Phys. Rev. Lett.} \textbf{\bibinfo{volume}{94}},
  \bibinfo{eid}{210601} (\bibinfo{year}{2005}).

\bibitem[{\citenamefont{F\`{e}ve et~al.}(2007)\citenamefont{F\`{e}ve, Mah\'{e},
  Berroir, Kontos, Placais, Glattli, Cavanna, Etienne, and Jin}}]{Feve07}
\bibinfo{author}{\bibfnamefont{G.}~\bibnamefont{F\`{e}ve}},
  \bibinfo{author}{\bibfnamefont{A.}~\bibnamefont{Mah\'{e}}},
  \bibinfo{author}{\bibfnamefont{J.-M.} \bibnamefont{Berroir}},
  \bibinfo{author}{\bibfnamefont{T.}~\bibnamefont{Kontos}},
  \bibinfo{author}{\bibfnamefont{B.}~\bibnamefont{Placais}},
  \bibinfo{author}{\bibfnamefont{D.}~\bibnamefont{Glattli}},
  \bibinfo{author}{\bibfnamefont{A.}~\bibnamefont{Cavanna}},
  \bibinfo{author}{\bibfnamefont{B.}~\bibnamefont{Etienne}}, \bibnamefont{and}
  \bibinfo{author}{\bibfnamefont{Y.}~\bibnamefont{Jin}},
  \bibinfo{journal}{Science} \textbf{\bibinfo{volume}{316}},
  \bibinfo{eid}{1169} (\bibinfo{year}{2007}).

\bibitem[{\citenamefont{Samuelsson et~al.}()\citenamefont{Samuelsson, Neder,
  and B\"uttiker}}]{Buttiker08}
\bibinfo{author}{\bibfnamefont{P.}~\bibnamefont{Samuelsson}},
  \bibinfo{author}{\bibfnamefont{I.}~\bibnamefont{Neder}}, \bibnamefont{and}
  \bibinfo{author}{\bibfnamefont{M.}~\bibnamefont{B\"uttiker}},
  \bibinfo{note}{cond-mat/0808.4090 (2008)}.

\bibitem[{\citenamefont{Levitov et~al.}(1996)\citenamefont{Levitov, Lee, and
  Lesovik}}]{LesovikLeeLevitov96}
\bibinfo{author}{\bibfnamefont{L.}~\bibnamefont{Levitov}},
  \bibinfo{author}{\bibfnamefont{H.}~\bibnamefont{Lee}}, \bibnamefont{and}
  \bibinfo{author}{\bibfnamefont{G.}~\bibnamefont{Lesovik}},
  \bibinfo{journal}{J.~Math.~Phys.} \textbf{\bibinfo{volume}{37}},
  \bibinfo{pages}{4845} (\bibinfo{year}{1996}),
  \bibinfo{note}{cond-mat/9607137}.

\bibitem[{\citenamefont{Moskalets and B\"{u}ttiker}(2007)}]{Moskalets07}
\bibinfo{author}{\bibfnamefont{M.}~\bibnamefont{Moskalets}} \bibnamefont{and}
  \bibinfo{author}{\bibfnamefont{M.}~\bibnamefont{B\"{u}ttiker}},
  \bibinfo{journal}{Phys. Rev. B} \textbf{\bibinfo{volume}{75}},
  \bibinfo{eid}{035315} (\bibinfo{year}{2007}).

\bibitem[{\citenamefont{Ivanov et~al.}(1997)\citenamefont{Ivanov, Lee, and
  Levitov}}]{IvaLL97}
\bibinfo{author}{\bibfnamefont{D.~A.} \bibnamefont{Ivanov}},
  \bibinfo{author}{\bibfnamefont{H.~W.} \bibnamefont{Lee}}, \bibnamefont{and}
  \bibinfo{author}{\bibfnamefont{L.~S.} \bibnamefont{Levitov}},
  \bibinfo{journal}{Phys. Rev. B} \textbf{\bibinfo{volume}{56}},
  \bibinfo{pages}{6839} (\bibinfo{year}{1997}).

\bibitem[{\citenamefont{Keeling et~al.}(2006)\citenamefont{Keeling, Klich, and
  Levitov}}]{KeelingKlichLev06}
\bibinfo{author}{\bibfnamefont{J.}~\bibnamefont{Keeling}},
  \bibinfo{author}{\bibfnamefont{I.}~\bibnamefont{Klich}}, \bibnamefont{and}
  \bibinfo{author}{\bibfnamefont{L.~S.} \bibnamefont{Levitov}},
  \bibinfo{journal}{Phys. Rev. Lett.} \textbf{\bibinfo{volume}{97}},
  \bibinfo{pages}{116403} (\bibinfo{year}{2006}).

\bibitem[{\citenamefont{Lee and Levitov}(1993)}]{LeeLevitov93}
\bibinfo{author}{\bibfnamefont{H.}~\bibnamefont{Lee}} \bibnamefont{and}
  \bibinfo{author}{\bibfnamefont{L.}~\bibnamefont{Levitov}}
  (\bibinfo{year}{1993}), \bibinfo{note}{cond-mat/9312013}.

\bibitem[{\citenamefont{Levitov and Lesovik}(1993)}]{LevitovJETP93}
\bibinfo{author}{\bibfnamefont{L.}~\bibnamefont{Levitov}} \bibnamefont{and}
  \bibinfo{author}{\bibfnamefont{G.}~\bibnamefont{Lesovik}},
  \bibinfo{journal}{JETP Letters} \textbf{\bibinfo{volume}{58}},
  \bibinfo{pages}{230} (\bibinfo{year}{1993}).

\bibitem[{\citenamefont{Muzykantskii and Adamov}(2003)}]{MA03}
\bibinfo{author}{\bibfnamefont{B.}~\bibnamefont{Muzykantskii}}
  \bibnamefont{and} \bibinfo{author}{\bibfnamefont{Y.}~\bibnamefont{Adamov}},
  \bibinfo{journal}{Phys. Rev. B} \textbf{\bibinfo{volume}{68}},
  \bibinfo{pages}{155304} (\bibinfo{year}{2003}).

\bibitem[{\citenamefont{d'Ambrumenil and Muzykantskii}(2005)}]{dAMuz05}
\bibinfo{author}{\bibfnamefont{N.}~\bibnamefont{d'Ambrumenil}}
  \bibnamefont{and}
  \bibinfo{author}{\bibfnamefont{B.}~\bibnamefont{Muzykantskii}},
  \bibinfo{journal}{Phys. Rev. B} \textbf{\bibinfo{volume}{71}},
  \bibinfo{pages}{045326} (\bibinfo{year}{2005}).

\bibitem[{\citenamefont{Vanevi\'{c} et~al.}(2007)\citenamefont{Vanevi\'{c},
  Nazarov, and Belzig}}]{VanevicNazBelz07}
\bibinfo{author}{\bibfnamefont{M.}~\bibnamefont{Vanevi\'{c}}},
  \bibinfo{author}{\bibfnamefont{Y.~V.} \bibnamefont{Nazarov}},
  \bibnamefont{and} \bibinfo{author}{\bibfnamefont{W.}~\bibnamefont{Belzig}},
  \bibinfo{journal}{Phys. Rev. Lett.} \textbf{\bibinfo{volume}{99}},
  \bibinfo{pages}{076601} (\bibinfo{year}{2007}).

\bibitem[{\citenamefont{Vanevi\'{c} et~al.}(2008)\citenamefont{Vanevi\'{c},
  Nazarov, and Belzig}}]{Nazarov08}
\bibinfo{author}{\bibfnamefont{M.}~\bibnamefont{Vanevi\'{c}}},
  \bibinfo{author}{\bibfnamefont{Y.~V.} \bibnamefont{Nazarov}},
  \bibnamefont{and} \bibinfo{author}{\bibfnamefont{W.}~\bibnamefont{Belzig}},
  \bibinfo{journal}{Phys.Rev. B} \textbf{\bibinfo{volume}{78}},
  \bibinfo{eid}{245308} (\bibinfo{year}{2008}).

\bibitem[{\citenamefont{Sherkunov et~al.}(2008)\citenamefont{Sherkunov, Pratap,
  Muzykantskii, and d'Ambrumenil}}]{sherkunov08}
\bibinfo{author}{\bibfnamefont{Y.~B.} \bibnamefont{Sherkunov}},
  \bibinfo{author}{\bibfnamefont{A.}~\bibnamefont{Pratap}},
  \bibinfo{author}{\bibfnamefont{B.}~\bibnamefont{Muzykantskii}},
  \bibnamefont{and}
  \bibinfo{author}{\bibfnamefont{N.}~\bibnamefont{d'Ambrumenil}},
  \bibinfo{journal}{Phys. Rev. Lett.} \textbf{\bibinfo{volume}{100}},
  \bibinfo{eid}{196601} (\bibinfo{year}{2008}).

\bibitem[{\citenamefont{Andreev and Kamenev}(2000)}]{Andreev-Kamenev2000}
\bibinfo{author}{\bibfnamefont{A.}~\bibnamefont{Andreev}} \bibnamefont{and}
  \bibinfo{author}{\bibfnamefont{A.}~\bibnamefont{Kamenev}},
  \bibinfo{journal}{Phys. Rev. Lett.} \textbf{\bibinfo{volume}{85}},
  \bibinfo{pages}{1294} (\bibinfo{year}{2000}).

\bibitem[{\citenamefont{Wiseman and Vaccaro}(2003)}]{Wiseman03}
\bibinfo{author}{\bibfnamefont{H.~M.} \bibnamefont{Wiseman}} \bibnamefont{and}
  \bibinfo{author}{\bibfnamefont{J.~A.} \bibnamefont{Vaccaro}},
  \bibinfo{journal}{Phys. Rev. Lett.} \textbf{\bibinfo{volume}{91}},
  \bibinfo{pages}{097902} (\bibinfo{year}{2003}).

\bibitem[{\citenamefont{Klich and Levitov}({\natexlab{a}})}]{Klich081}
\bibinfo{author}{\bibfnamefont{I.}~\bibnamefont{Klich}} \bibnamefont{and}
  \bibinfo{author}{\bibfnamefont{L.}~\bibnamefont{Levitov}},
  \bibinfo{note}{quant-ph/0804.1377 (2008)}.

\bibitem[{\citenamefont{Klich and Levitov}({\natexlab{b}})}]{Klich082}
\bibinfo{author}{\bibfnamefont{I.}~\bibnamefont{Klich}} \bibnamefont{and}
  \bibinfo{author}{\bibfnamefont{L.}~\bibnamefont{Levitov}},
  \bibinfo{note}{quant-ph/0812.0006 (2008)}.

\bibitem[{\citenamefont{Abanov and Ivanov}(2008)}]{Abanov08}
\bibinfo{author}{\bibfnamefont{A.~G.} \bibnamefont{Abanov}} \bibnamefont{and}
  \bibinfo{author}{\bibfnamefont{D.~A.} \bibnamefont{Ivanov}},
  \bibinfo{journal}{Phys. Rev. Lett.} \textbf{\bibinfo{volume}{100}},
  \bibinfo{eid}{086602} (\bibinfo{year}{2008}).

\end{thebibliography}

\end{document}